\documentclass[11pt,a4paper]{article}
\usepackage{jheppub}

\def\p{\partial}
\def\beq{\begin{eqnarray}}
\def\eea{\end{eqnarray}}
\def\be{\begin{equation}}
\def\ee{\end{equation}}
\begin{document}

\title{On the Hagedorn behavior of the superstring propating in a cosmological time dependent background}


\author{Daniel Luiz Nedel}

\affiliation{Universidade Federal da Integra\c{c}\~ao Latino-Americana \\
Avenida Tancredo Neves 6731, Foz do Igua\c{c}u, Brasil}

\emailAdd{daniel.nedel@unila.edu.br}

\abstract{In this work the LvN quantization of the type IIB superstring is carried on in a time dependent plane wave background with a constant self-dual Ramond-Ramond 5-form  and a linear dilaton in the light-like direction.  Such an endeavour allows us to define an invariant density matrix and study important issues in real time string thermodynamics. In particular, the Hagendorn temperature is calculated as function of the thermalization time. }


\arxivnumber{0000.0000}


%
\maketitle

\flushbottom

\section{Introducion}
The formulation of superstring theory at finite temperature  and the study of the superstring sigma model for time dependent backgrounds are topics of constant and renewed interest in the literature, in view of the structural role  superstring theory plays in the construction of theoretic frameworks for fundamental interactions and quantum gravitation. In particular , the study of thermal effects  when the string propagates in  cosmological time dependent geometries can shed  light on important questions related to quantum cosmology and  may help to understand the nature of space-like
singularities\cite{Berkooz:2007nm}.

One outstanding 
feature of string theory at finite temperature is the exponential growth of states as function
of energy. Due to this behavior, the partition function becomes ill defined for temperatures above the so-called Hagedorn temperature.
If the Hagedorn behavior works in string theory as it works in hadrons physics, then the
true degrees of freedom of the theory at high temperature may be others than those of the
perturbative string. However, in spite of many works about finite temperature string theory,
a precise understanding of the Hagedorn temperature and the true degrees of freedom at
higher temperatures is still lacking.  Many of the advances made in understanding the Hagedorn temperature stem from a specific equilibrium finite temperature field theory formalism: the imaginary time formalism. In this case,  the thermal state is described
by compactifying Euclidean time on a circle, the thermal circle . The radius of the thermal circle is equal to the inverse temperature in natural  units. For string theory applications this formalism entails two complications. The first one arises from the simple fact that string theory contains gravity: for theories containing gravity, the radius
of the thermal circle becomes a dynamic field, which makes the very notion of thermal equilibrium non-trivial \cite{Gross:1982cv}.   In addiction, for closed strings one needs to take into account the winding modes around the thermal circle .  Above the Hagedorn temperature these
modes become tachyonic and it is precisely these tachyonic excitations that encodes the Hagedorn divergence and the long/short string transition discussed in \cite{Barbon:2001di}, \cite{Brustein:2022uft}, \cite{Salomonson:1988ac},\cite{Salomonson:1985eq}. However, when the superstring propagates in a time dependent background, the mass and coupling parameters of the superstring sigma model
depend explicitly on time. Therefore, the study of thermal effects in this time dependent  superstring sigma model  requires a real time formalism.  Actually, from a worldsheet perspective it's  an open system. So, a non equilibrium formalism must be taken into account.   

In general, the non equilibrium quantization of a determined system is carried on using the  Schwinger and Keldysh formalism.
 In this formalism a closed time path integral is introduced  to treat properly the
  non equilibrium evolution of quantum fields from their initial thermal equilibrium. Here another approach is used: it is the so called 
  Liouville-von Neumann (LvN) approach \cite{Kim,KMMS,Lewis}. 

The LvN approach
is a canonical method that unifies the usual methodology  to study the evolution of pure states, given by the functional Schrodinger equation, with the usual approach used to study the evolution of mixed states, described by the density matrix (which in turn obeys the LvN equation). Note that even though the density matrix depends on time, it  still satisfies the LvN equation. Hence, the LvN  method treats the time-dependent
nonequilibrium system exactly in the same way as the time-independent one.

In the present work the LvN approach is used to study thermal effects in the  light cone superstring for the superstring propagating in a time dependent plane wave background with a constant self-dual Ramond-Ramond 5-form and a linear dilaton in the light-like direction. 
This background keeps sixteen
supersymmetries and the sigma model was canonically
quantized  in \cite{Bin}, where it was shown that the Hamiltonian is time-dependent with vanishing zero-point energy and has
a supersymmetric spectrum. As shown in \cite{Bin}  the background  is geodesically incomplete and hence admits a null cosmology interpretation. However, the dilaton diverges close to the cosmological singularity and so one needs a nonperturbative description to study the  string dynamics close to null singularity.
In the sigma model studied in \cite{Bin}, the sign that the theory is not well defined at the singularity appears as a divergence in the time-dependent Hamiltonian when it is evaluated close to the singularity.
 On the other hand, it was shown in \cite{Marchioro:2020qub}  that  as the string evolves  towards the singularity, the vacuum  seen by  asymptotically flat observers is a left/right  entanglement state. Hence a left/right superstring entanglement entropy appears, dynamically generated by the background. It was shown that, at the singularity, the left/right string entanglement is finite an then could be a useful tool to probe the singularity.  
Furthermore, it was shown that, at the singularity, the left/right entanglement state is in fact a thermal state and the worldsheet entanglement entropy becomes the thermodynamic entropy for a 2d free supersymmetric gas, which implies that near the singularity the string thermalizes at a finite temperature.   Here, in order to study  more carefully the superstring thermalization in this background, the superstring canonical quantization   is carried on in the Liouville picture, 
 which allows us to calculate the Hagedorn temperature in the adiabatic approximation as function of time, where time means the time where thermalization takes place. In fact, the Hagedorn temperature is shown to increase as the string evolves from the  asymptotically flat time to the singularity  time.  The present work is divided as follows: in the first section the LvN approach is presented. The time dependent background studied here is presented in section \ref{rev}. In section \ref{LN} the bosonic and fermionic sectors of the light cone superstring time dependent sigma model are quantized in the LvN picture and the invariant  creation/annihilation operators are constructed. The density matrix and the adiabatic approximation are discussed in section \ref{den}. As an application, the non equilibrium  thermal two point function  is calculated in the adiabatic approximation. Finally in section \ref{Hag} the Hagedorn temperature is calculated as function of time.

\section{The  Liouville-von Neumann (LvN) method}

The core of the LvN approach lies on the definition of  invariant operators and the fact that quantum LvN equation provides all the quantum and statistical information of non equilibrium systems.  Given an  operator O and a  time dependent evolution operator $U(t)$ , an invariant operator $O_L(t)$ is defined by $O_L(t)= U(t)O_SU^{\dagger}(t)$, where $O_S$ is the operator $O$ in  Schrodinger picture. This relation also defines the so called Liouville picture. Comparing to the Heisenberg picture,  the  operator $O_L(t)$ evolves backward in the same way as the density operator. So $O_L$ also satisfies the  LvN equation
\begin{equation}
    i\frac{\partial O_L}{\partial t}+\left [O_L,H\right]=0 \label{LvN}
\end{equation}
where the Hamiltonian H can be time dependent. The Lewis-Riesenfeld invariant theorem states that an operator satisfying (\ref{LvN}) has time-dependent eigenstates and time-independent eigenvalues. So the spectrum of the invariant operators yields  quantum states for the time dependent system.

In order to study the nonequilibrium evolution  exactly in the same way as the equilibrium one, it is necessary to find an invariant operator $O_L$ such that a time dependent density matrix satisfying LvN equation can be written as $\rho_L(t)= Z^{-1}e^{-\beta O_L(t)}$, where Z is the trace of  $\rho_L(t)$.  Here it is assumed that the system reaches a thermodynamic equilibrium point characterized by the temperature $1/\beta$. At the time $t_0$ at which equilibrium is reached, $\rho_L(t_0)$ is the usual equilibrium density matrix. As an example, 
suppose we have an oscillator interacting with a thermal bath and, as a consequence of this interaction, we have a time-dependent mass. The Hamiltonian will be time dependent and there will be  modes creation(or particle creation in a quantum field scenario). One could naively construct a thermal density matrix defined by the time dependent Hamiltonian

\begin{equation}
\rho_H=\frac{1}{Z}e^{-\beta H(t)} \,.
\label{rhoh}
\end{equation}

\noindent This density matrix does not satisfy the quantum Liouville-von Neumann (LvN) equation and it is not possible to relate $1/\beta$ to the equilibrium temperature. If the system starts in the initial thermal equilibrium state, its final state can be far away from the initial one. Actually , owing to particle production, the final state can be  unitarily inequivalent to the initial one.
The strategy of the LvN approach is to define time dependent oscillators $a_L,a_L^{\dagger}$ that satisfy the equation (\ref{LvN}). 
\begin{equation}
i\frac{\partial a_L}{\partial \tau} +[a_L,H] =0 \,. \label{osinv}
\end{equation} 
The linearity of the LvN equation allows us to use ${a}_L(t)$ and
${a}_L^{\dagger}(t)$ to construct operators that also satisfy equation (\ref{LvN}); in particular, the number operator $N_L={a_L}_L^{\dagger} (t) {a_L} (t)$ . By using the Lewis-Riesenfeld invariant theorem, one finds the Fock space consisting of the time dependent number
states   such that
\begin{equation}
{N}_L (t) \vert n, t \rangle = n \vert n, t\rangle \,.
\label{numst}
\end{equation}

\noindent With the invariant oscilators, a density matrix which satisfies the LvN equation can be defined as

\begin{equation}
\rho_{\rm T} = \frac{1}{Z_N}e^{\beta\omega_0 a^{\dagger}(t)a(t)} \,,
\label{rhoT}
\end{equation}

\noindent where $\beta$ and $\omega_0$ are free parameters and the trace that appears in the definition of $Z_N$ is taken over the states defined in (\ref{numst}). Now, the system is characterized  by the density matrix (\ref{rhoT}) in the same way of a time independent one. The key point is to find solutions for equation (\ref{osinv}) such that, in the adiabatic regime, the density matrix (\ref{rhoT}) is equal to the density matrix (\ref{rhoh})  evaluated at the thermalization time.  Thus, the unitary real time evolution of the system can be studied until it reaches thermal equilibrium, characterized by $\beta$.

\section{ The Background} \label{rev}

In this section, the time dependent background studied here  is presented.  
Consider the following time dependent background with Ramond-Ramond flux

\begin{eqnarray}
ds^2 = -2dx^+ dx^- -\lambda(x^+)\,x_I^2\,dx^+ dx^+ +dx^Idx^I\,,\nonumber \\
\phi=\phi(x^+)\,,\quad\quad(F_5)_{+1234}=(F_5)_{+5678}=2f,
\label{BG}
\end{eqnarray}

\noindent where   $\phi$ is the dilaton and $F_5$ the Ramond-Ramond field.  As usual for a generic plane wave, the supersymmetry preserved by the background is reduced from maximal (32 supercharges) to sixteen supercharges. When type IIB Green-Schwarz (GS) superstring propagates in this background, conformal invariance of the worldsheet demands
 
\begin{eqnarray}
R_{\mu\nu}=-2D_{\mu}D_{\nu}\phi+\frac{1}{24}e^{2\phi}(F^2_{5})_{\mu\nu}\,,
\label{Conformal}
\end{eqnarray}
\noindent and the only non zero component of the Ricci curvature tensor $R_{\mu\nu}$  is
\begin{equation}
R_{++}=8\lambda(x^+).
\end{equation}
Putting (\ref{BG}) into (\ref{Conformal})  gives

\begin{eqnarray}
\lambda=-\frac{1}{4}\phi''+f^2e^{2\phi}\,.
\end{eqnarray}
\noindent In the reference \cite{Bin}, a solution of (\ref{Conformal}) with non zero constant Ramond-Ramond field ($f=f_0$) is studied. It has the form

\begin{equation}
\phi= -c x^+,\:\: \lambda= f_0^2e^{-2cx^+}, \label{modelo}
\end{equation}

\noindent for any constant $c$. In this case, the metric  admits a null cosmology interpretation and  the cosmological singularity is located in $x^+= -\infty$. Note that  in this model the string coupling $g= e^{-\phi}$ diverges at the singularity.   As discussed previously, interaction of the string with this kind of backround makes the parameters of the sigma model time-dependent. In the next sections the quantization of the superstring sigma model for this background is carried on  in the Liouville picture.

\section{Superstring in LvN picture.} \label{LN}
In this section the  Liouville-von Neumann method is used to study the quantum dynamics of the superstring propagating in the backgound  (\ref{BG}). This implies defining a superstring Hilbert space constructed with creation/annihilation operators that are LvN invariant, that is, operators which satisfy equation (\ref{LvN}).  
\subsection{Bosonic Sector}
Let us start  with the bosonic sector. Although the gauge fixing  has already been discussed in \cite{Bin}, it is useful to include a
review here in order to fix notation. The bosonic part of the superstring sigma model  for the background (\ref{BG}) is
\beq
&&S=\frac{1}{4\pi\alpha'}\int d^2\sigma g^{ab}\ G_{\mu\nu}\partial_aX^\mu\partial_bX^\nu \nonumber \\
&&=\frac{1}{4\pi\alpha'}\int d^2\sigma g^{ab}\ \left(-2\partial_aX^+\partial_bX^- 
+\partial_aX^I\partial_bX^I -m^2(X^+) X_I^2\partial_aX^+\partial_bX^+\right) , \nonumber \\
\eea
where $g_{ab}$ is the worldsheet metric, $\sigma^a=(\tau, \sigma)$ are the worldsheet coordinates, $I=1,2,\cdots ,8$ and $m(X^+)= fe^{-cX^+} $. As usual,
 the RR  fluxes do not appear in the bosonic action. The  bosonic worldsheet gauge symmetry is fixed using the light cone gauge
\beq\label{gauge1}
\sqrt{-g}g^{ab}&=&\eta^{ab}\ ,\ \ \ \ -\eta_{\tau\tau}=\eta_{\sigma\sigma}=1\ \nonumber \\
X^+& =&\alpha' p^+\tau\ ,\ \ \ \  p^+>0\ .
\eea

In this gauge all the dynamics is determined by $X^I$'s through the constraints resulting from
\cite{Green:2012oqa}
\be
\frac{\delta {\cal L}}{\delta g_{\tau\sigma}}=0 \ , \ \ \
\frac{\delta {\cal L}}{\delta g_{\tau\tau}}=\frac{\delta {\cal L}}{\delta g_{\sigma\sigma}}=0 \label{constr}
\ee
After setting $-g_{\tau\tau}=g_{\sigma\sigma}=1$, the constraints (\ref{constr})  allow us to write $\partial_{\sigma}X^-$ and  $\partial_{\tau}X^-$in terms of $X^I$

\be\label{Virasoro}
\partial_{\sigma}X^-=\frac{1}{\alpha' p^+} \ \partial_{\sigma}X^I\partial_{\tau}X^I \ , 
\ee
\be\label{lightconeX-}
\partial_{\tau}X^-=\frac{1}{2\alpha' p^+} \ \biggl(
\partial_{\tau}X^I\partial_{\tau}X^I+\partial_{\sigma}X^I\partial_{\sigma}X^I-
(m(\tau)\alpha' p^+)^2 X^IX^I\biggr) \, .
\ee
Choosing $c= \frac{1}{\alpha^{\prime}p^+}$, the light cone bosonic action can be written as
\be\label{LCbosonicaction}
S^{bos.}_{l.c.}=\frac{1}{4\pi\alpha'}\int d\tau\int_0^{2\pi\alpha' p^+} d\sigma  \left[
\partial_\tau X^I\partial_\tau X^I -\partial_\sigma X^I\partial_\sigma X^I 
-m^2(\tau) X_I^2\right] \, ,
\ee
where $\tau$ and $\sigma$ were re-scaled by $\alpha' p^+$ and $m(\tau)=f_0e^{-\tau}$.
Since the bosonic sector of the theory is $SO(8)$ invariant, the $I$ index   will be
frequently omitted.

In order to quantize the theory in the LvN approach,  the string coordinate $X(\sigma)$ and momentum density
$P(\sigma)= \frac{\dot{X}}{2\pi\alpha^\prime} $ are expanded as 

\beq
X^I(\sigma)& = &x_{0}^I+\sqrt{2}\sum_{n=1}^{\infty}
\left(x_{n}^I\cos\frac{n\sigma}{\alpha}+x_{-n}^I\sin\frac{n\sigma}{\alpha}\right)\,,\nonumber \\
P^I(\sigma) & =&\frac{1}{2\pi\alpha}\left[p_{0}^I+\sqrt{2}\sum_{n=1}^{\infty}
\left(p_{n}^I\cos\frac{n\sigma}{\alpha}+p_{-n}^I\sin\frac{n\sigma}{\alpha}\right)\right]\,, \label{bosoniexp}
\eea
where it was defined $\alpha=p^+\alpha^\prime$. In this notation(usual in pp waves backgrounds) all of the string oscillations—left-movers, right-movers, and the zero
modes—can be treated on an equal footing.  Note that the form of the expansion for the worldsheet fields $X^I(\sigma)$ and $P^I(\sigma)$ allows us to
 associate the  Fourier modes  to Hermitian operators $x_n$ and $p_n$. This will be very useful for writing the thermal density matrix in the position representation.
In general the Fourier mode operators $x_n$ and $p_n$ can be time dependent. However, in the LvN picture they are Schr\"{o}dinger operators.  They can be chosen such that the expansion (\ref {bosoniexp}) represents $X$ and $P$ at a given fixed time.  

The normalization is chosen so that the canonical
commutation relation
\beq
[X^I(\sigma), P^J(\sigma')] = i \delta^{IJ}\delta(\sigma-\sigma')
\eea
follows from imposing
\be
[{x}_m^I, {p}_n^J] = i\delta^{IJ}\delta_{mn}.
\ee
Next,  the light cone Hamiltonian is written as 
\be\label{bosoniH}
H^{bos.}_{l.c.}=\frac{1}{4\pi\alpha'}\int_0^{2\pi\alpha' p^+}d\sigma \ 
\big[(2\pi\alpha')^2 P_I^2 +(\partial_\sigma X^I)^2+m^2(\tau) X^2_I \big] \, .
\ee
In order to proceed with the LvN quantization,  equation 
(\ref {bosoniexp}) is used to write the light cone Hamiltonian in terms of the Fourier mode operators( omitting $SO(8)$ indices):

\be
H^{bos.}_{l.c.} = \frac{1} {2\alpha} \sum_{n=-\infty}^\infty
\left[ {p}_n^2 +  \omega_n^2(\tau) {x}_n^2 \right],
\ee
where $\omega_n(\tau)=\sqrt{n^2+ \alpha^2m^2(\tau)}$.
Now, the LvN invariant operators can be can be found. Following the LvN approach, it can be defined a set of bosonic  rising and lowering operators
\be
\left [\alpha_n^I(\tau),\alpha_m^{\dagger\,J}(\tau)\right]=\delta^{IJ}\delta_{nm}
\ee \label{boscilator}
satisfying the quantum light cone
LvN equation 
\beq
\frac{i}{\alpha^{\prime}p^+}\frac {\partial }{\partial \tau} \alpha_n^I(\tau) + [\alpha_n^I(\tau),H^{bos.}_{l.c.}]=0 , n\in \mathbb{Z}. \label{lnvb}
\eea
In order to find the invariant bosonic string oscillators, the operators  $\alpha_n(\tau)$,$\alpha_m^{\dagger}(\tau)$ are defined in terms of the Fourier mode operators $x_n$, $p_n$
\beq
\alpha_n^I(\tau) &=& i\left(\phi_n^*(\tau)p_n^I -\dot{\phi_n^ *}(\tau){x_n^I}\right) \nonumber \\
\alpha_n^{\dagger\,I} (\tau) &=& -i\left(\phi_n(\tau)p_n^I  -\dot\phi_n(\tau){x_n^I}\right )\, ,\label{invb} 
\eea
where $\phi$ and $\dot \phi$ must satisfy the Wronskian
\be
{\dot\phi}^*_n(\tau)\phi_m(\tau) -\dot\phi(\tau)\phi^{*}(\tau)=i\delta_{mn} \label{wronkb}
\ee
to ensure that the relations (\ref{boscilator}) are satisfied. Now all work boils down to finding the functions  $\phi_n(t)$ such that $\alpha_n^I(\tau)$ satisfy (\ref{lnvb}). Plugging (\ref {invb}) into (\ref {lnvb}) results in the following equation for $\phi_n(t)$
\be
\ddot\phi_n + \omega_n^2(\tau)\phi_n=0\,.
\ee
A solution that satisfies (\ref{wronkb}) can be written  in terms of Bessel Functions:
\begin{equation}
\phi_n(\tau)=\sqrt{ \left(\frac{\tilde{f}}{2}\right)}\Gamma(1+in)J_{in}\left(z(\tau)\right)\:,
\end{equation}
where $\tilde{f}=\alpha f_0$,  $z(\tau)=\tilde{ f}e^{-\tau}$ and $J_m$ is a Bessel function of the first kind. The relations (\ref{wronkb})  follow from the Gamma and  Bessel function properties
\begin{eqnarray}
&&\Gamma(1+in)\,\Gamma(1-in)=\frac{n\pi}{\,\sinh{n\pi}}\,\,,\nonumber\\
&&J_{\nu}(z)J'_{-\nu}(z)-J_{-\nu}(z)J'_{\nu}(z)=-\frac{2\sin{\nu\pi}}{\pi z}\,\,.
\label{Bes}
\end{eqnarray}
 Here we are interested in the adiabatic limit given by 
\begin{equation}
    |\frac{\dot{\omega}_n(\tau)}{\omega(\tau)}|=\tilde{f}^2\frac{\tilde{f}^2e^{-2\tau}}{\tilde{f}^2e^{-2\tau}+n^2}\ll 1
\end{equation}
Note that, even close to the null singularity ($\tau\rightarrow -\infty$), the adiabatic regime is controlled by Ramond-Ramond field. So, in the adiabatic regime ($\alpha f_0<<1$), the solution can be approximated  by
\begin{equation}
\phi_n (\tau) \approx  \phi_n^{ad} (\tau)=\frac{1}{\sqrt{2 \omega_n (\tau)}} e^{- i \int \omega_n
(\tau)},
\end{equation}
The adiabatic solution is important to study the thermalization process in the LvN approach. It will be shown that in this regime the invariant density matrix approaches the thermal equilibrium density matrix, calculated at the instant of time when the system enters thermodynamic equilibrium.  Once the invariant creation and annihilation operators are defined, one uses the Lewis-Riesenfeld theorem and finds a base  for the bosonic  fock space defined by  time-dependent number states:
\be
N_n^{LvN}(\tau)|n,\tau\rangle_b=n|n,\tau\rangle_b
\ee
where $N_n^{LvN}(\tau)$ is defined in the usual way: 
\be
N_n^{LvN}(\tau)= \delta^{IJ}\alpha_n^{\dagger I}(\tau)\alpha_n^J(\tau). \label{Ninva}
\ee
 In the next section, the invariant number operator will be used to define a LvN invariant density matrix.  Suppose the system thermalizes adiabatically at time $\tau_0$. Then, close to $\tau_0 $ the Hamiltonian  can be written as
\begin{equation}
H^{bos.}_{l.c.}\approx H^{bos.}_{LvN}= \frac{1 }{ \alpha}\left[\sum_{n=-\infty}^\infty \omega_n (\tau_0)N_n^{LvN}(\tau) + 4\right]. \label{Hinv}
\end{equation}

The position and momentum modes operators that appear in equation (\ref{bosoniexp} )can be also defined  in the Heisenberg representation. Let us define  ${p}_n(\tau)$ and  ${x}_n(\tau)$ as the  momentum and position modes operators in the Heisenberg representation, and write the non invariant operators


\beq  
a_n^I(\tau)&=& \frac{\sqrt{2\omega_n(\tau)}}{2}\left[\frac{p_n^I}{\omega_n}-ix_n^I \right]\nonumber \\
a_n^{\dagger\, I}(\tau)&=& \frac{\sqrt{2\omega_n(\tau)}}{2}\left[\frac{p_n^I}{\omega_n}+ix_n^I\right]\, ,  \label{timea}
\eea
  which obey
\be
[ a_m^I(\tau), a_n^{\dagger\,J}(\tau)] = \delta_{mn}\delta^{IJ}.
\ee
In terms of the non-invariant creation and annihilation operators,  the time dependent bosonic Hamiltonian may be written as
\be
H^{bos.}_{l.c.}(\tau)  = \frac{1 }{ \alpha}\left[\sum_{n=-\infty}^\infty \omega_n (\tau)a_n^\dagger(\tau)
a_n(\tau) + 4\right] \label{Hb}\, ,
\ee
which is identical to the bosonic part of the Hamiltonian derived in \cite{Bin}. The Hamiltonian is diagonal; however, it cannot be used to define the thermal density matrix since it is not LvN invariant.

\subsection{Fermionic Sector}
In this subsection the quantization of the fermionic sector is worked out in the LvN approach. As in the bosonic case, for the sake of self-containedness a review of the light cone gauge fixing process, without further details, will be carried out.  The fermionic part of the type two B superstring in this background can be written as
\begin{eqnarray}
S^{fer.}=-\frac{i}{2\pi\alpha'}\int d^2\sigma
(\sqrt{-g}g^{ab}\delta_{AB}-\epsilon^{ab}\sigma_{3AB})\,\partial_{a}x^{\mu}\,
\bar{\theta}^A\Gamma_{\mu}(\hat{D}_b\theta)^B + {\mathcal O} (\theta^3) \,,
\label{FerA}
\end{eqnarray}

\begin{eqnarray}
\sigma_3 = \mbox{diag}(1,-1)\:, \nonumber\\
\hat{D}_b=\partial_b+\Omega_\nu\,\partial_{b}x^\nu \:,
\end{eqnarray}

\noindent with $\hat{D}_b$ being the pull-back of the covariant derivative to the worldsheet. The indices $a,b$ are worldsheet indices; $A,B = 1,2$ and $\mu$ is the spacetime index. The spin connection $\Omega_\nu$ is defined by

 \begin{eqnarray}
\Omega_-&=&0,\nonumber\\
\Omega\,_I&=&\frac{i\,e^\phi}{4}f\,\Gamma^+(\Pi+\Pi')\,\Gamma_I\,\sigma_2,\nonumber\\
\Omega_+&=&-\frac{1}{2}\lambda\,x^I\Gamma^{+I}\textbf{1}+
\frac{i\,e^\phi}{4}f\,\Gamma^+(\Pi+\Pi')\,\Gamma_+\sigma_2 \:,
\label{CovDiff}
\end{eqnarray}
 where $\Gamma^{\pm} = (\Gamma^0 \pm \Gamma^9)/\sqrt{2}$, $\sigma_2$ is the Pauli matrix and $\Pi$ is symmetric, traceless and squares to one \footnote{ The following representation will be used : $\Pi = \Gamma^{1}\Gamma^{2}\Gamma^{3}\Gamma^{4} = \mbox{diag} ({\mathbf 1}_4, -{\mathbf 1}_4)$, $\Pi' =  \Gamma^{5}\Gamma^{6}\Gamma^{7}\Gamma^{8}$}. The fermionic fields $\theta^A$ are 10d spinors and in equation (\ref{FerA})  the space time spinor indices where omitted( actually  $\theta^A=\theta^A_\alpha$  with $\alpha=1,2,\ldots,16\,$, and $A=1,2$). Higher orders in theta will not be taken into account because they do not contribute in the light-cone gauge \cite{Metsaev:2001bj},\cite{Sadri}. The representation of $\Gamma$-matrices chosen is such that $\Gamma^0$ is the 10d charge conjugation; therefore, the components of $\theta^{A}$ are all real.  The  gauge symmetries are fixed  choosing light-cone gauge:
 
\begin{eqnarray}
 x^+ = \alpha' p^+ \tau \:, \:\:\: p^+ > 0 \:. \nonumber\\ 
\Gamma^+ \theta^A = 0 \:,
\label{gfix}
\end{eqnarray} 
The kappa symmetry ($\Gamma^+ \theta^A = 0 $) implies
\begin{eqnarray}
(\theta^A)^T \Gamma^I \theta^B = 0, \:\: \forall A, B \:, \nonumber\\
(\Omega_I)^{A}_{\:\:B} \theta^B = 0 \:, \nonumber\\
\Pi \theta^A = \Pi' \theta^A \:.
\label{kappa}
\end{eqnarray}
After fixing the kappa symmetry  the ten dimensional fermions are reduced to $SO(8)$ representation.  In ref.\cite{Bin} the light cone fermionic action is written in terms of the real fields $\theta^1_{a}$ and $\theta^2_{a}$; here  complex fields will be used.  Since  $\theta^1_{a}$ and $\theta^2_{a}$  have the same chirality, we can define
the  complex positive chirality  SO(8) spinor ($\theta^a$,  $a=1,\ldots,8$) by
\begin{equation}
\theta_a=e^{-i\frac{\pi}{4}}\left(\theta^1_{a}+i\theta^2_{a}\right),\:\:\:\bar{\theta}_a=e^{i\frac{\pi}{4}}\left(\theta^1_{a}-i\theta^2_{a}\right)
\end{equation}
 From this point onwards the $SO(8)$ spinor indices will  be often omitted. Finally, the light cone fermionic action can be written as 
\begin{equation}
S^{fer.}_{l.c.} = \frac{1}{  4 \pi\alpha'} \int d\tau \int_0^{2 \pi \alpha} d\sigma\ [i(\bar{\theta}
\p_\tau \theta + \theta \p_\tau \bar{\theta}) - \theta \p_\sigma \theta
+ \bar{\theta} \p_\sigma \bar{\theta} - 2
m(\tau) \bar{\theta} \Pi \theta].
\end{equation}
where $\tau$ and $\sigma$ were re-scaled as in the bosonic case . The last term in the action is a time dependent mass term resulting from the RR five-form flux. The time dependent mass $m(\tau)$ is the same as the bosonic sector.
Again, in order to quantize the theory using the LvN approach, we expand
$\theta$ and its conjugate momentum 
$\Lambda \equiv \frac{i} { 2 \pi\alpha'}
\bar{\theta}$ as
\begin{eqnarray}
\theta(\sigma)&= &\vartheta_0 + \frac{1} { \sqrt{2}}
\sum_{n \ne 0} ( \vartheta_{|n|} - i e(n) \vartheta_{-|n|})
e^{i n \sigma/\alpha'p^+},\cr
\Lambda(\sigma) &=& 
\frac{i}{2 \pi \alpha} \left[ \lambda_0 + \frac{1} {\sqrt{2}}
\sum_{n \ne 0} (\lambda_{|n|} - i e(n) \lambda_{-|n|})
e^{i n \sigma/\alpha'p^+}\right], \label{expf}
\end{eqnarray}
such that the anticommutation relation
\begin{equation}
\{ \theta^a(\sigma), \Lambda^b(\sigma') \} =
i\delta^{ab}
\delta(\sigma - \sigma')
\end{equation}
follows from
\begin{equation}
\{ {\vartheta}_m^a,
{\lambda}_n^b \} = \delta^{a b} \delta_{mn}.
\end{equation}
In equation (\ref{expf}), $e(n)$ is the signal of $n$. Note that the  Fourier modes satisfy $\lambda_n =\frac{\alpha'p^{+}}{2}\bar{\vartheta_n}$.
For the sake of simplicity, let us define $\lambda=-i\Lambda$. In terms of $\lambda$, the  fermionic Hamiltonian is
\begin{equation}
H^{fer.}_{l.c.} = 
 \frac {1}{2}\int_0^{2 \pi \alpha'p^{+}}
d\sigma\ \left[-4 \pi  \lambda \p_\sigma
\lambda + \frac{1} {4 \pi} \theta \p_\sigma \theta + 2  m(\tau)
(\lambda \Pi \theta)\right].
\end{equation}
In terms of the Fourier mode operators,
we have
\begin{equation}
H^{fer.}_{l.c.} =\frac{1}{2} \sum_{n=-\infty}^\infty\left[\frac{n}
{ 2}\left(\frac{4}{ (\alpha^{\prime}p^{+})2} {\lambda}_{-n}
{\lambda}_n - \hat{\vartheta}_{-n}
{\vartheta}_n\right)
+2 m(\tau) {\lambda}_n \Pi {\vartheta}_n
\right].
\end{equation}

If this Hamiltonian is used to solve the LvN equation in order to find the invariant fermionic operators, a set of coupled equations that are difficult to solve will emerge. The equations become simpler if we perform the following Bogoliubov transformation
\begin{eqnarray}
    \lambda_n&=&\frac{\sqrt{\alpha}}{2}\left[\hat{\lambda}_n+e(-n)\hat{\vartheta}_{-n} \right], \:\:n\neq 0 \nonumber \\
    {\vartheta}_n&=&\frac{1}{\sqrt{\alpha}}\left[\hat{\vartheta}_n +e(-n)\hat{\lambda}_{-n}\right], \:\:n\neq 0 \nonumber \\
    \lambda_0&=&\hat{\lambda}_0,\:\:\: {\vartheta}_0=\hat{\vartheta}_0,
\end{eqnarray}
 such that 
\begin{equation}
    \{\hat{\lambda}_n, \hat{\vartheta}_m\}=\delta_{nm},\:\: n\in \mathbb{Z}
\end{equation}
In terms of the hat operators, the Hamiltonian is written as
\begin{equation}
H^{fer.}_{l.c.}= H^{fer.}_0+\sum_{n=1}^{\infty}\left[\frac{n}{\alpha}\left(  \hat{\vartheta}_{-n}\hat{\lambda}_{-n}-\hat{\lambda}_n\hat{\vartheta}_n \right) +m(\tau)\left(\hat{\lambda}_{-n}\Pi\lambda_n +\hat{\vartheta}_n\Pi\hat{\vartheta}_{-n}\right)\right],
\end{equation}
where
\begin{equation}
    H^{fer.}_0= \tilde{f}^2e^{-2\tau}\hat{\lambda}_0\Pi\hat{\vartheta}_0.
\end{equation}

Now, it can be defined a set of fermionic LvN invariant operators satisfying
\begin{equation}
\{ \beta_m(\tau), \beta_n^\dagger (\tau)\} = \delta_{mn}\, , n\in \mathbb{Z}\, . \label{invosf}
\end{equation}

Let's write $\beta_n(\tau)$ and $\beta_n^\dagger (\tau)$ as
\begin{eqnarray}
    \beta_n(\tau)= F(\tau)\hat{\lambda}_n+G(\tau)\hat\vartheta_{-n} \nonumber \\
    \beta_n^{\dagger}(\tau)= F(\tau)^*\hat{\vartheta}_n+G(\tau)^*\hat\lambda_{-n} ,
\end{eqnarray}
where the functions $F(\tau)$ and $G(\tau)$ must satisfy
\begin{equation}
|F(\tau)|^2+|G(\tau)|^2=1 \label{fercond}.
\end{equation}
The equations (\ref{LvN}) for $\beta_n$ result in the 
following system of coupled first order equations 
\begin{eqnarray}
i\dot{F}+nF+\alpha m(\tau)\Pi G&=&0 \nonumber \\
i\dot{F}-nG+ \alpha m(\tau) \Pi F&=&0.
 \label{coupled}
\end{eqnarray}
By using $\Pi^2=1$, equations (\ref{coupled})  results in the following
decoupled second order equations :

\begin{eqnarray}
   \ddot{G}+\dot{G}+(n^2+in+\tilde{f}^2e^{-2\tau})G&=&0 \nonumber \\
   \ddot{F}+\dot{F}+(n^2-in+ \tilde{f}^2e^{-2\tau})F&=&0
    \end{eqnarray}

By defining again $z(\tau)= \tilde{f}e^{-\tau}$, the solutions that satisfy the conditions (\ref{fercond}) are
\begin{eqnarray}
F(\tau)&=& \sqrt{\frac{z(\tau)}{2}}\Gamma(\frac{1}{2}+in)J_{\frac{1}{2}+in}\left(z(\tau)\right) \nonumber \\
G(\tau) &=& \sqrt{\frac{z(\tau)}{2}}\Gamma(\frac{1}{2}+in)J_{-\frac{1}{2}+in}\left(z(\tau)\right),
\end{eqnarray}
where it was used the following properties
\begin{eqnarray}
&&\Gamma \left(\frac{1}{2}+in \right)\,\Gamma\left(\frac{1}{2}-in \right)=\frac{\pi}{\,\cosh{n\pi}}\,\,,\label{GaHalf}\\
&&J_{-\frac{1}{2}+in}(z)J_{-\frac{1}{2}-in}(z)+J_{\frac{1}{2}+in}(z)J_{\frac{1}{2}-in}(z)
=\frac{2\cosh{n\pi}}{\pi z}.\label{J-Hal}
\end{eqnarray}

 An adiabatic solution can be found and it has the same structure as that found in the bosonic case. So, we have constructed a set of fermionic rising and lowering
invariant operators that can be used to  define the fermionic number operators in the usual way. The superstring vacuum is defined by
\begin{equation}
    \alpha_n^I(\tau)|0,t\rangle=0,\:\:\:\:\beta_n(\tau)|0,\tau\rangle=0
\end{equation}
The states created from $\alpha_n^{\dagger},\beta_n^{\dagger}$ in general depend on time, but  the Lewis-Riesenfeld theorem guarantees that their eigenvalues (occupation
numbers) do not depend on time .

As in the bosonic case, it can also be found a set of  non-invariant time dependent fermionic rising and lowering/
operators, satisfying
\begin{equation}
\{ b_m(\tau), b_n^\dagger(\tau) \} = \delta_{mn}, n\in \mathbb{Z}.
\end{equation}
 However, change of basis is far more complicated than the one used in the bososic sector. Going back to base (\ref{expf}), let's write  ${\vartheta}_{n}$ and ${\lambda}_{n}$ as the fermionic mode operators in the Heisenberg representation. It can be defined the following time dependent operators:
\begin{eqnarray}
b_n(\tau) &=& \frac{1}{2}\left[ \sqrt{\alpha}A^{+}_n(\tau){\vartheta}_{n}+ \frac{2}{\sqrt{\alpha}}e(n)A^-_n(\tau)\lambda_{-n}\right] \nonumber \\
b_n^{\dagger}(\tau) &=& \frac{1}{2}\left[ \frac{2}{\sqrt{\alpha}}A^{+}_n(\tau){\lambda}_{n}+ \sqrt{\alpha}e(n)A^-_n(\tau)\vartheta_{-n}\right], \label{noinvafer}
\end{eqnarray}
where   the time dependent matrices $A^{\pm}_n(\tau)$ are defined by
\begin{equation}
A^{\pm}_n(\tau)= \frac{1}{1+ \gamma_n(\tau)}\left(1+\gamma_n(\tau)\Pi\right),\:\: \gamma_n(\tau)=\frac{\omega_n(\tau) - |n|}{ \alpha m(\tau)}.
\end{equation}

The change of basis is similar to the one used in the time independent case  \cite{Spradlin:2002ar}. Note that the relations (\ref {noinvafer}) also breaks the $SO(8)$ symmetry to $SO(4)\times SO(4)$.  The matrices $A^{\pm}_n(\tau)$ were chosen in such a way that in terms of  $b_n$, the fermionic time dependent Hamiltonian(non invariant) is diagonal and takes the simple form
\begin{equation}
H^{ferm.}_{l.c.} (\tau)=
\frac{1}{ \alpha'p^+}
\left[
\sum_{n=-\infty}^\infty \omega_n(\tau) \left(b_n^\dagger b_n 
- 4\right) \right], \label{ferH}
\end{equation}
such that the total time dependent  superstring Hamiltonian, written in terms of $a_n$, $b_n$, has the same form of the one found in \cite{Bin} using a different method.  Note that the zero-point energy of the non invariant Hamiltonian exactly cancels between the bosons and the fermions. 

\section{The invariant string density matrix} \label{den}
 In the previous section the superstring was quantized in LvN picture, which made it possible to find  LvN invariant creation/annihilation operators. These  invariant operators can be used to define a density matrix that satisfies the  light cone LvN equation:
 \begin{equation}
\frac{i}{\alpha} \frac{\partial}{\partial \tau}\rho_{LvN}+\left[\rho_{LvN},H_{l.c.}(\tau)\right] \label{invd}
 \end{equation}
 where $H_{l.c.}= H_{l.c.}^{bos.}+ H_{l.c.}^{fer.}$. Note that, in terms of the non invariant oscillators, the Hamiltonian is:
 \begin{equation}
     H_{l.c.} (\tau)= \frac{1}{\alpha}\sum_{n=-\infty}^\infty \omega_n(\tau) \left( a_n^\dagger(\tau) a_n(\tau)  +b_n^\dagger(\tau) b_n(\tau)\right).  \label{timeH}
 \end{equation}
  This Hamiltonian is diagonal and has a time dependent supersymetric spectrum.  However, as explained before, this Hamiltonian cannot be used to define a thermal density matrix because it is not LvN invariant. The main goal of this section is to show that, if the thermalization occurs adiabatically at time $\tau_0$, this Hamiltonian can be used to define an instantaneous thermal density matrix, defined at time $\tau_0$. To this end, it will be defined first a density matrix that satisfies equation (\ref{invd})  an then will be shown that, in the adiabatic limit, this density matrix approaches the one obtained with the instantaneous  Hamiltonian $H_{l.c.} (\tau_0)$. With this result, we can calculate the thermal partition function at an equilibrium temperature T, defined in $\tau_0$ in terms  of the instantaneous diagonal Hamiltonian. As an application of the  invariant density matrix, the non equilibrium worldsheet two point function is calculated. For the sake of simplicity, only the bosonic sector are going to be taken into account. The generalization for the fermionic sector is straightforward. 
 
In order to calculate the invariant light cone density matrix , we need to take into account the time like killing vectors. In  flat space, the time like killing vector  is $\frac{1}{\sqrt{2}}\left[\frac{\partial}{\partial x^+}+\frac{\partial}{\partial x^-}\right]$. Here, owing to the non trivial dilaton dependency  on $x^+$ , the timelike killing vector is just $\frac{\partial}{\partial x^-}$. However, we want to define an invariant density matrix such that, in the asymptotically flat limit, it reduces to the standard expression for the light cone flat space string density matrix. So, the invariant  density matrix is defined as

\be
\rho_{LvN}= \frac{1}{Z_{LvN}} e^{-\tilde{\beta}\left(p^+ +H_{LvN}\right)}\,,
\ee
where $\tilde{\beta}= \frac{\beta}{\sqrt{2}}$ and  $H_{LvN}$ is given in terms of the invariant creation/annihilation  operators
\be 
H_{LvN}= = \frac{1}{\alpha'p_+}\sum_{n=1}^\infty\tilde{\omega}_n\left(\delta_{IJ}\alpha^{I\dagger}_n\alpha^{J}_n+\delta_{ab}\beta^{a\dagger}_n\beta^{b}_n\right).
\ee
The normalization factor $Z_{LvN}$(the string partition function) is the trace 
\be
Z_{LvN}=Tr e^{-\tilde{\beta}\left(p^+ + H_{LvN}\right)}.
\ee
  In general, $\tilde{\omega}_n$ and $\beta$ are free parameters.  It is  assumed that the system thermalizes at a time $\tau_0$, with equilibrium temperature $1/\beta$. The  parameter $\tilde{\omega}_n$ will be related to $\omega_n(\tau_0)$, as it will be clear soon .
 
 By using the Lewis-Riesenfeld invariant theorem, the Hamiltonian can be written on a time dependent number basis and the density matrix can be written as
\begin{eqnarray}
\rho_{LvN}(\beta,\tau)
=\frac{1}{Z_{LvN}}\sum_{\{ {n_i^I}\}}  \exp{-\tilde{\beta}\left[\displaystyle\sum_{I,i}\tilde{\omega}_in_i^I+ p^+\right]}|\{ {n_i^I}\} ,p^+,\tau\rangle\langle \{n_i^I\},p^+,\tau|, \nonumber \\ \label{rholvn}
\end{eqnarray}
where $\{n_i^I\}=\{n_i^I\}^\infty_{i=-\infty}=n_{-\infty}^I,\dots,n_\infty^I$. 
In order to simplify the notation, space-time indices will not be taken into account in the next steps.  As the background is symmetrical with respect to the different transverse coordinates,  one can calculate the contribution of one dimension and then take into account the other transverse dimensions. 
Now let us take advantage of the notation used in (\ref {bosoniexp}) to write the density matrix in position representation. In terms of the Fourier modes it takes the form
\be
\rho^1(x_1,x_2,...,x^\prime_1,x^\prime_2...,\beta)_{LvN}=  \frac{1}{Z_{LvN}}\langle x_1,x_2,...|\rho_{LvN}(\tilde{\beta})|x_1^\prime,x_2^\prime,...\rangle,
\ee
where index 1 in the density matrix indicates that the contribution of only one  transversal dimension is being taken into account. To simplify the notation, $\rho^1(x,x^\prime,\beta)_{LvN}$ will be used instead of $\rho^1(x_1,x_2,...,x^\prime_1,x^\prime_2...,\beta)_{LvN}$. The next step is to write  the string number state in position representation. By writing $\alpha_n(\tau)$ in the position representation, the LvN vacuum  state is defined by
\be
i\left[\phi^*_n\frac{\partial}{i\partial x_n}-\dot{\phi}_nx_n\right]\Psi_0=0,\:\:\: n\in \mathbb{Z}\,.
\ee
The normalized solution is
\be
\Psi_0=\prod_{j\in \mathbb{Z}}\left(\frac{1}{2\phi_j}\right)^{1/4}e^{\frac{i}{2}
\frac{\dot{\phi_j}^*}{\phi_j}x_j^2}\,.
\ee
The other states are constructed in the usual way by applying $\alpha_n^{\dagger}(\tau)$. So, the string number state is written in the  coordinate representation as
\be
\Psi_n =\prod_j\frac{1}{\sqrt{2\pi\phi^*_j\phi_j}}\frac{1}{\sqrt{2^{n_j}n_j!}}\left(\frac{\phi_j}{\phi^*_j}\right)H_{n}(q_j)e^{\frac{i}{2}\frac{\dot{\phi}^*_j}{\phi_j}x_j^2}  \label{hermit}\,,
\ee
where $H_{n}(q_j)$ are the Hermite polynomials and  
\be
q_j= \frac{x_j}{\sqrt{2\phi_j^*\phi_j}}\,.
\ee
Using (\ref{hermit}), the density matrix for one transversal coordinate $\rho^1(x,x^\prime,\beta)$ is given by
\be
\frac{e^{-\tilde{\beta} p^+}}{Z_{LvN}}\sum_{\{n_j\}}\prod_{j\in\mathbb{Z}}\frac{H_{n_j}(q_j)H_{n_j}(q_j^\prime)}{2\pi\phi_j^*\phi_j2^{n_j}n_j!}e^{-\tilde{\beta}\omega_j(n_j+\frac{1}{2})}
e^{i[\frac{\dot{\phi_j}^*}{\phi_j}x_j^2- \frac{\dot{\phi_j}}{\phi_j^*}{x_j^\prime}^2]}\,.\\
\ee
Following the method developed in \cite{Kubo}, the density matrix can be simplified using the following integral representation for the Hermite polynomials:
\be
H_{n_j}(q_j)=\frac{1}{\sqrt{\pi}}\int_{-\infty}^{\infty}(-2iz)^{n_j}e^{-(z_j+iq_j)^2}dz.
\ee
 Using this identity for each mode, one has
\be 
\rho^1(x,x^\prime,\beta)= \frac{ e^{-\tilde{\beta} p^+} }{Z}\sum_{n_j}\prod_{j\in\mathbb{Z}}\frac{1}{\sqrt{2\pi\phi_j^*\phi_j}}e^{-\tilde{\beta} \frac{\omega_j}{2}}
I_{n_j}\:,
\ee
where 
\be I_{n_j} = \frac{e^{-\tilde{\beta}\omega_jn_j}}{\pi2^{n_j}n_j!}e^{q_j^2+{q_j^\prime}^2}\int\int dz_jdw_j(2iz_j)^{n_j}(2iw_j)^{n_j}e^{-z_j^2+2iz_jx_j}e^{-w_j^2+2iw_jx_j^{\prime}}.
\ee
Now, by defining the following matrices
\beq 
A_j= 2\begin{bmatrix} 1&e^{-\tilde{\beta}\omega_j} \\
e^{-\tilde{\beta}\omega_j} &1 
\end{bmatrix}, \:\: Y_j=\begin{bmatrix} z_j\\w_j\end{bmatrix},\:\:B_j=-2\begin{bmatrix}x_j^\prime\\x_j\end{bmatrix},
\eea
after summing over each $n_j$, the density matrix is 
\be 
\rho^1(x_1,x_2,...,x^\prime_1,x^\prime_2...,\beta)_{LvN}=\frac{e^{-\tilde{\beta} p^+}}{Z}\sum_{n_j}\prod_{j\in\mathbb{Z}}\frac{1}{\sqrt{2\pi^2\phi_j^*\phi_j}}e^{q_j^2+{q_j^\prime}^2}\int\int exp\left[-\frac{1}{2}Y^\dagger {\bf A_j}Y_j +i B_j^\dagger Y_j\right].
\ee
Finally, one can use the result
\beq 
\int\int e^{\left[ -\frac{1}{2} Y^\dagger {\bf A}Y +i B^\dagger Y\right]}= \frac{2\pi}{\sqrt{\det {\bf A}}}e^{\left[\frac{1}{2}B^\dagger{\bf A}^{-1}B \right]}
\eea 
to  write the density matrix in the form( after some algebra and taking into account the eight transverse dimensions) 
\begin{eqnarray}
\rho(x_1,x_2,...,x^\prime_1,x^\prime_2...,\beta)_{LvN} &=& \frac{e^{-\beta p^+}}{Z_{LvN}}\prod_{j\in \mathbb{Z}} \Biggl[\frac{1}{2 \pi
 \phi_j^*(\tau) \phi_j(\tau)\sinh\tilde{\beta}\tilde{\omega}_j}\Biggr]^{4}\nonumber \\
 &\times&\exp
\Biggl[ 4i\sum_{n\in \mathbb{Z}} \frac{d}{d\tau} \ln ( \phi_j^*(\tau) \phi_j(\tau)) ({x'_n}^2 - x_n^2)
\Biggr] \nonumber\\ &\times& \exp\Biggl[-\sum_{n\in \mathbb{Z}}\frac{1}{\phi(\tau)_j^* \phi(\tau)_j } \Biggl\{
(x_n' + x_n)^2 \tanh(\frac{\tilde{\beta}\tilde{\omega}_j}{2}) + (x_n'-x_n)^2
\coth(\frac{\tilde{\beta} \hbar \tilde{\omega}_j}{2}) \Biggr\} \Biggr].\nonumber \\  \label{rhox}
\end{eqnarray}
One can now compare this density matrix with the density matrix obtained with the  time dependent Hamiltonian  (\ref{timeH}) defined at a time $\tau_0$
\be \rho_T(\tau_0)= \frac{e^{-\tilde{\beta}\left(p^++ H^{bos}_{lc}(\tau_0)\right)}}{Z_T(\tau_0)},
\ee
where $T= \frac{1}{\beta}$ is the equilibrium temperature and $Z_T$ is the thermal partition function
\be Z_T(\tau_0) = Tr e^{-\tilde{\beta}\left(p^++ H^{bos}_{lc}(\tau_0)\right)}\,. \label{partition}
\ee

It is easy to see that when $\tilde{\omega}_n =\omega_n(\tau_0)$ , $Z_T=Z_{LvN}$. Following the same steps as before, the instantaneous density matrix is written in the position representation as
\beq
\rho_T(\tau_0) &=& \frac{e^{-\tilde{\beta} p^+}}{Z_T(\tau_0)}\prod_{j\in \mathbb{Z}} \Biggl[\frac{\omega_j(\tau_0)}{2 \pi
  \sinh( \tilde{\beta} \omega_j(\tau_0)}\Biggr]^{4}  \nonumber\\ 
  && \times \exp\Biggl[-2\sum_{n\in \mathbb{Z}}\omega_n(\tau_0)\Biggl\{
(x_n' + x_n)^2 \tanh(\frac{\tilde{\beta}  \omega_n(\tau_0)}{2}) + (x_n'-x_n)^2
\coth(\frac{\tilde{\beta}  \hbar \omega_n(\tau_0)}{2}) \Biggr\} \Biggr]. \nonumber \\ \label{thermalden}
\eea
In the adiabatic regime
\be \phi_n\phi^*_n \approx \frac{1}{2\omega_n},\:\:\:\left|\frac{\dot{\omega_n}}{\omega_n}\right|<<1,
\ee
one has
\be
\rho(x_1,x_2,...,x^\prime_1,x^\prime_2...,\beta)_{LvN}\approx \rho_T(\tau_0),
\ee
if one sets $\tilde{\omega}_j =\omega_j(\tau_0)$. So, close to $\tau_0$, the non equilibrium string thermal state is given by  (\ref{rholvn}) and (\ref{rhox}) an it can be used (\ref{partition}) to calculated, for example,  the Hagedorn temperature as a function of $\tau_0$. Before, as a direct application of the invariant density matrix, let`s calculate the worldsheet time dependent  two-point function at finite temperature, which is an important object to study non-equilibrium phenomena.
\subsection{The Real Time thermal two point function.}

As an application of the  invariant density matrix, it can be calculated the time dependent  two-point function at finite temperature. Let's start using the LvN invariant operators to evaluated the two-point function at equal times  at zero temperature, by taking the expectation
value with respect to the vacuum state  $|0,\tau\rangle$ which is annihilated by $\alpha_n^I(\tau)$,
\begin{equation} 
g^{IJ}(\sigma,\sigma^{\prime})= \langle 0,\tau|X^I(\sigma,\tau)X^J(\sigma^{\prime},\tau)|0,\tau\rangle \,.
\end{equation}
This is computed by inverting the relations (\ref{invb} ):
\begin{equation}
    x_n^I=\alpha_n^I(\tau)\phi_n(\tau)+\alpha_n^I(\tau)\phi_n^*(\tau).
\end{equation}

In the adiabatic approximation one gets
\begin{equation} 
g^{IJ}(\sigma,\sigma^{\prime},\tau)= \alpha^\prime\sum_{n\in \mathbb{Z}}\frac{e^{i(\sigma-\sigma^\prime)}}{\omega_n(\tau)}.
\end{equation}

  The time dependent finite temperature two-point function is calculated  by taking the expectation
value with respect to the thermal state, defined by the invariant density matrix
\be 
G^{IJ}(\sigma,\sigma^{\prime},t)_T= \mathrm{Tr}\left[\rho_{LvN}X^I(\sigma,\tau)X^J(\sigma^\prime,\tau)\right ].
\ee
The trace must be taken over the physical states of the closed string. Although the lightcone gauge solves the two worldsheet reparametrization constraints, one single consistency condition remains to be imposed to calculate the trace above,  which is related to the circle isometry of the closed string. In order to fix this isometry on the Fock space, the bosonic physical states $|\Psi\rangle_b$ must be annihilated by the  $\sigma$ translation generator ${\cal P}_b$, which implies the level-matching constraint
 \be {\cal P}_b|\Psi\rangle_b=0\ee
 where 
 \be
 {\cal P}_b = \sum_{n\in \mathbb Z}n\left[\delta_{IJ}\alpha_n^ {I\dagger }(\tau)\alpha_n^J(\tau) \right].
 \ee 
 Then, to ensure that the trace  is taken over the physical states, one introduces the projector
 \be \int_{-1/2}^{1/2} d\lambda e^{\left(2\pi i\lambda {\cal P}_b\right)},
 \ee
 such that the invariant thermal two point function is
\be
G^{IJ}(\sigma,\sigma^{\prime},\tau)_T=\frac{1}{Z_{LvN}} \int dp^{+} e^{-\tilde{\beta} p^{+}}\int_{-1/2}^{1/2}\sum_{\{n_j\}}\langle \{n_j(\tau)\}|e^{-\tilde{\beta}H_{LvN} }e^{2\pi i\lambda {\cal P}}X^I(\sigma,\tau)X^J(\sigma^\prime,\tau)|\{n_j(\tau)\}\rangle,
\ee
where $H_{LvN}$ is given in (\ref{Hinv}). 
By defining
\be  k_j = \tilde{\beta}\omega_j(\tau_0), \:\:\: b_j =2\pi \lambda j ,
\ee
the two point function can be written as
\be G^{IJ}(\sigma,\sigma^{\prime},\tau)_T= \frac{1}{Z}\int p^+\sum_{\{n_j\}}\langle n_j|\exp{\left[-\displaystyle\sum_{j\in \mathbb{Z}}(k_j+ib_j)n_j\right ]}X(\sigma,\tau)X(\sigma^\prime,\tau)|\{n_j\}\rangle.
\ee 

After some algebra, the two point function is written as a zero temperature two point function plus a thermal contribution 
\beq G^{IJ}(\sigma,\sigma^{\prime},\tau)_T = \frac{\alpha^{\prime}}{Z}\int dp^+\int_{-1/2}^{1/2} d\lambda |\eta_m(\beta,\lambda)|^8
\left[g^{IJ}(\sigma,\sigma^{\prime},\tau)+ 2\alpha^\prime \delta^{IJ} g(\sigma,\sigma^{\prime},t)_T\right]\nonumber \\
\eea
where $\eta_m(\beta,\lambda)$ is the "massive" eta function
\be  \eta_m(\beta,\lambda) = \prod_{n\in \mathbb{N}}\frac{1}{1-e^{-\tilde{\beta}\omega_n(\tau_0)+2\pi i\lambda n}} \ee
and 
\beq  g(\sigma,\sigma^{\prime},\tau)_T &=& \sum_{n=0} \frac{1}{\omega_n}\left[\frac{e^{-(k_n-ib_n)}}{1-e^{-(k_n-ib_n)}}\cos n\sigma\cos n\sigma^\prime +
\frac{e^{-(k_n+ib_n)}}{1-e^{-(k_n+ib_n)}}\sin n\sigma\sin n\sigma^\prime \right] \nonumber \\
&=&\sum_{p=1,n=0}\frac{e^{-pkn}}{\omega_n}\left[e^{ipb_n}
 \cos n\sigma cos n\sigma^\prime +e^{-ipb_n}\sin n\sigma\sin n\sigma^\prime \right]
\eea
is the thermal correction. In general, in finite temperature quantum field theories the term in $\eta$ in the numerator does not appear because it is just the Z factor of the denominator. Here these terms are left over due to the integral over $p^+$ and $\lambda$(note that $\omega_n$ depends on $p^+$). 

Le`s focus on $g(\sigma,\sigma^{\prime},t)_T$.
The parity of the eta function in relation to $\lambda$ can be used to rewrite the thermal contribution to two point function as
\beq
 g(\sigma,\sigma^{\prime},\tau)_T&=&  \sum_{p=1}^{\infty}\sum_{n\in \mathbb{Z}}\frac{e^{-pK_n}}{2\omega_n}\left[e^{i n\left(2\pi p\lambda + (\sigma-\sigma^\prime)\right) } + e^{i n\left(2\pi p\lambda - (\sigma-\sigma^\prime)\right)  }\right ].
 \eea
  In order to investigate the leading short-distance  behaviour , the Poisson resummation formula  is used:
  \be 
  \sum_{n\in \mathbb{Z}}F(n) = \sum_{l\in \mathbb{Z}}\int_{-\infty}^{\infty}e^{2\pi iyl}F(y)dy ,\label{poisson}
  \ee 
along with the following representation of the modified Bessel function,
  \beq 
  K_0 &=& \int_0^\infty\frac{e^{-\beta\sqrt{x^2 +\gamma^2}}}{\sqrt{x^2 +\gamma^2}}\cos b x dx= K_0(\sqrt{b^2+ \beta^2}),
 \label{k0}
\eea 
to rewrite the   finite temperature two point function as
\beq 
G^{IJ}(\sigma,\sigma^{\prime},\tau) &=& 2\alpha^{\prime}K_0(fe^{-\tau}|\sigma-\sigma^{\prime}|) \nonumber \\
                                  &+&  \frac{2 \alpha^{\prime}}{Z}\int dp^+\int_{-1/2}^{1/2} d\lambda |\eta_m(\beta,\lambda)|^8\sum_{n\in \mathbb{Z}} K_0(fe^{-\tau}|2\pi n \alpha+\sigma- \sigma^\prime| ) \nonumber \\
                                  &+& \frac{2\alpha^{\prime}}{Z}\int dp^+\int_{-1/2}^{1/2} d\lambda |\eta_m(\beta,\lambda)|^8\sum_{p=1}^{\infty}\sum_{n\in \mathbb{Z}}\left[K_0(fe^{-\tau}\sqrt{(\beta\alpha p)^2+ b_{n}^+(\sigma,\sigma^\prime,\lambda)}\right]\nonumber  \\
                                  &+& \frac{2 \alpha^{\prime}}{Z}\int dp^+\int_{-1/2}^{1/2} d\lambda |\eta_m(\beta,\lambda)|^8\sum_{p=1}^{\infty}\sum_{n\in \mathbb{Z}}\left[K_0( fe^{-\tau}\sqrt{(\beta \alpha p)^2+ b_{n}^-(\sigma,\sigma^\prime,\lambda)}\right],\nonumber  \\
                                  \eea
where  $b_n^{\pm}(\sigma,\sigma^\prime,\lambda)=\left[2\pi\alpha (n+\lambda)\pm (\sigma-\sigma^\prime)\right]^2 $. We can see that the only term that has singularities  when $\sigma \rightarrow \sigma^{\prime}$ is the first term. In particular,  the finite temperature contributions is finite at short-distances.  Let`s analyze the behaviour of the two point function in two different limits. In the limit $fe^{\tau}|\sigma -\sigma^\prime|<<1$, one can expand the Bessel function as
\be 
K_{0}\left(z\right)=-\left(\ln\left(\tfrac{1}{2}z\right)+\gamma\right)I_{0}%
\left(z\right)+\frac{\tfrac{1}{4}z^{2}}{(1!)^{2}}+(1+\tfrac{1}{2})\frac{(%
\tfrac{1}{4}z^{2})^{2}}{(2!)^{2}}+(1+\tfrac{1}{2}+\tfrac{1}{3})\frac{(\tfrac{1%
}{4}z^{2})^{3}}{(3!)^{2}}+\dotsi,
\ee 
where $I_{0}(z)$  is the modified Bessel function of first kind ($I_{0}(0 )= 1$) and $\gamma $ is the Euler constant.  So,  the leading short-distance
behavior of the  two point function is
\be  \alpha^{\prime}\ln \frac{ fe^{-\tau}}{|\sigma-\sigma^\prime|}
\ee 
which has the same leading short-distance logarithmic behavior of the flat space one. On the other hand, in the limit
$e^{-\tau}|\sigma-\sigma^\prime| >>1$, the Bessel function has the following asymptotic expansion
\be
K_0(z)= \sqrt{\frac{\pi}{2x}}e^{-x}\sum_{k=0}^{\infty}\frac{\Gamma(k+1/2)}{k!\Gamma(1/2-k)}(2z)^{-k} \,.
\ee 
In this limit, the thermal two point function has an exponential damping behavior. In particular, the two point function goes to zero near the null singularity. This may corroborate the idea  raised in  \cite{Madhu}, where it was argued  that the string gets highly excited and breaks up
into bits propagating independently near the singularity.
\section{ The  ligh cone Superstring Partition Function  and Hagedorn Behavior.} \label{Hag}
In this section the light cone superstring partition function will  be calculated. As previously shown, in the adiabatic regime the density matrix $\rho_T(\tau)$ \ constructed with the non-invariant operators approaches the invariant density matrix as $\tau$ approaches $\tau_0$, where $\tau_0$ is the time at which $\rho_T(\tau)$ is evaluated. This result allows us to use the diagonal Hamiltonian (\ref{timeH}) evaluated at $\tau_0$ to calculate the partition function and, consequently, the Hagedorn temperature as a function of $\tau_0$.  
 
 The  light cone superstring partition function at time $\tau_0$ is the trace
\beq
Z(\beta,\tau_0) &=& \mathrm{Tr}e^{-\tilde{\beta}  \left(p^++ H_{l.c.}(\tau_0)\right)}.
\eea
Again, in order to fix  the $S^1$ isometry on the superstring Fock space and to ensure  that the trace  is taken over the physical states, one introduces the projector
  \be
\int d\lambda e^{\left(2\pi i\lambda {\cal P}\right)},
 \ee 
 where the superstring sigma  translation generator ${\cal P}$  is
 \be
 {\cal P} = \sum_{n\in \mathbb Z}n\left[\delta_{IJ}a_n^ {I\dagger }a_n^J + \delta_{ab}b_n^{\dagger a}b_n^b\right]
 \ee 
and  ($a_n^{I\dagger}$,  $a_n^J$, $b_n^{\dagger a}$,$b_n^b$) are the operators (\ref{timea}) defined at time $\tau_0$.
The light cone superstring   thermal partition function can be written as
 \be Z(\beta,\tau_0) = \int dp^+\int d\lambda e^{-\beta p^+}\mathrm{Tr}e^{\left(-\tilde{\beta}  H_{l.c.}(\tau_0)+2i\pi \lambda {\cal P}\right)}.
 \ee
The integrand of this partition function has interesting modular properties,  which become apparent by defining the following complex parameter
\be
\tau^{\prime} = \lambda +i\frac{\tilde{\beta} }{2\pi\alpha^\prime p^+} =\tau_1+i\tau_2, \label{moduli}
\ee 
such that the partition function can be written as

 \be Z(\beta,\tau_0) =\int dp^+ e^{-\beta p^+}\mathrm{Tr}\left[e^{-2\pi \tau_2 H_{l.c.(\tau_0)}}e^{i2\pi\tau_1{\cal P} }\right].
\ee 
It is well known that the thermal partition function of the closed close string can be written as a functional integral on the torus \cite{Polchinski:1985zf}. Let's remember here how the torus appears in the density matrix formalism that we are using. Note that the operator $e^{-2\pi\tau_2H_{l.c.}}$  propagates the closed superstring  through  imaginary light cone time $-2\pi\tau_2$. In turn, the operator $e^{i2\pi\tau_1 {\cal P}}$ rotates the closed string by an angle $2\pi\tau_1$. So, the trace taken over matrix elements of the form $\langle i|e^{-2\pi \tau_2 H_{l.c.(\tau_0)}}e^{i2\pi\tau_1{\cal P}}|f\rangle $ can be  represented as a path integral  on a torus by gluing the ends of the cylinder of length $2\pi \tau_2$ with a relative twist $2\pi\tau_1$. Actually  the  twist  is related to the Dehn twist associated to one of the cycles \cite{Castellani:1991ev}. We then conclude that $\tau^{\prime}$ is indeed the modulus of a torus represented by the  parallelogram defined in the complex plane with vertices at $0$, $\tau^{\prime}$, $1$, $\tau^{\prime}+1 $ and identified opposite sides. Furthermore, the thermal density matrix allows observing a kind of torus generalization of the KMS condition that is a consequence of the closed string torus topology \cite{Abdalla:2004dg}.

In order to explore the torus modular properties of the partition function, the integral over $p^+$ is rewritten as an integral over the moduli $\tau_2$, such that the UV asymptotic behavior of the partition function is recover in the  limit $\tau_2\rightarrow 0$. Finally, after  taking the trace over the bosonic and fermionic number states, the partition function is
\beq  
Z(\beta, \tau_0)=\frac{\tilde{\beta}}{ 2\pi \alpha^\prime}  
\int_0^\infty \frac{d\tau_2}{\tau_2{}^2}\int d\tau_1  
\exp(-\frac{\beta^2}{2\pi\,\alpha'\,\tau_2}) z_{lc}(\tau^{\prime},\tau_0 ),  
\label{Z}
  \eea  
where 
\be z_{lc}(\tau^{\prime},\tau_0 )= z_{lc}^{bos.}(\tau^{\prime},\tau_0 )z_{lc}^{ferm.}(\tau^{\prime},\tau_0) \ee
is the product of  the bosonic and fermionic contributions:
\beq  
z_{lc}^{bos.}(\tau^{\prime},\tau_0 )&=&\exp\left[  
-16\pi\tau_2\left(\frac {\tilde{f}e^{-\tau_0}}{2}+\sum_{n=1}^\infty \sqrt{n^2+\tilde{f}^2e^{-2\tau_0}}\right)\right]  
\nonumber\\  
&&\left[\prod_{n\in \mathbb{Z}}\left(1-\exp[2\pi (-\tau_2 \sqrt{n^2+\tilde{f}^2e^{-2\tau_0}}+i\tau_1 n)]  
\right)\right]^{-8},\label{zb}  
\eea  
\beq 
z_{lc}^{ferm.}(\tau^{\prime},\tau_0) &=&\exp\left[  
16\pi \tau_2\left(\frac {\tilde{f}e^{-\tau_0}}{2}+\sum_{n=1}^\infty \sqrt{n^2+\tilde{f}^2e^{-2\tau_0}}\right)  
\right]\nonumber\\  
&&\left[ 
\prod_{n\in{\bf Z}}\left(1+\exp[2\pi (-\tau_2 \sqrt{n^2+\tilde{f}^2e^{-2\tau_0}}  
+i\tau_1 n)]\right)\right]^8.\nonumber\\  
\label{zf}  
\eea  
 As usual in pp waves, the partition function is written in terms of generalized " massive"  modular functions. Note that the  contribution from the Ramond field now depends on the  torus moduli space from $p^+$, so  
 \begin{equation}
 \tilde{f}= \frac{\tilde{\beta}}{2\pi \tau_2}f_0.
 \end{equation}
 This does not happen in the  time dependent pp wave model studied in \cite{Blau:2004cm}, hence the partition function UV behavior of the two models are completely different. Actually, owing to scale  invariance of the metric, the  partition function  of the model studied in \cite{Blau:2004cm}  has the same UV behavior of the  string partition function in flat space.

We can now study the behavior of the partition function for each time $\tau_0$ where the thermalization occurs. In particular, it can be studied the UV behavior for each $\tau_0$. Before, let's assume that the thermalization occurs close to the null singularity and try to extrapolate this result to the strong coupling region. As it can be seen, there are no divergences in the partition function (added to those that we should have in the UV limit) due to singularity. 
This can be easily proven by performing a sequence of steps, which will also be useful for analyzing the UV behavior. Let's start
 by taking the logarithm of $z_{lc} (\tau^{\prime},\tau_0 )$
 
\beq  \ln z_{lc} (\tau^{\prime},\tau_0 ) =\nonumber 
\eea
\beq
8\sum_{n\in{\bf Z}} \left[ \log(1-e^{-2\pi\tau_2\sqrt{m^2+n^2}
+2\pi i\tau_1(n)+2\pi ia})
-\log(1+e^{-2\pi\tau_2\sqrt{m^2+(n)^2}+2\pi i\tau_1(n-1/2)
}\right] \nonumber \\
=-\sum_{n\in{\bf Z}}\sum_{p=1}^{\infty}\frac{1}{p}\left[ e^{-2\pi p\tau_2\left(-2\pi\tau_2\sqrt{m^2+n^2} \right)}F(n,p,\tau_1)\right]\nonumber \\ \label{lnz}
\eea
where \be
F(n,p,\tau_1) = 8e^{i2\pi n p}(1-\cos \pi p). \ee
Next, by making the replacement $r=p^2 s$ and  using the identity 
\be
e^{-z} = \frac{1}{ \sqrt{\pi}} \int_0^\infty dr\, r^{-1/2}
e^{-r-\frac{z^2}{ 4r}} \,,
\label{identi}
\ee
 equation ($\ref{lnz}$) becomes
\beq
\ln z_{lc} (\tau,\tau_0 )&=& \frac{1}{ \sqrt{\pi}} \sum_{n\in \mathbb{Z}}
\sum_{p=1}^{\infty}
\int_0^{\infty} ds\, s^{-1/2} e^{-p^2 s -  (2\pi\tau_2)^2\left( f^2e^{-2\tau_0}+n^2\right)/s}F(n,p,\tau_1) \nonumber \\
&=& 2\sum_{n\in \mathbb{Z}}
\sum_{p=1}^{\infty} \frac{(f^2e^{-2\tau_0}+n^2)^{1/4}}{\sqrt{\pi p}} K_{1/2}\left(2p\sqrt{f^2e^{-2\tau_0}+n^2}\right )\label{fstep}\, ,
\eea 
where it was used the following integral representation of the modified Bessel function:
\be 
K_{\nu}= \int_0^{\infty}s^{\nu-1}e^{-\frac{a}{s}-\frac{b}{s}}= 2\left(\frac{a}{b}\right)^{\nu/2}K_{\nu}\left(2\sqrt{ab}\right)\,. \label{besselint}
\ee
  Now, just by using the asymptotic behavior of the Bessel function 
  \be 
  \lim_{x\rightarrow \infty} K_{\nu}(x) \approx \sqrt{\frac{\pi}{2x}}e^{-x}\, , \label{besselassy}
  \ee 
  it  can be easily seen that there are no divergences in the partition function arising from the singular behavior of the metric in $\tau_0 \rightarrow -\infty$. This is not surprising since the partition function of genus one does not depend on string coupling.
 
 Next, the UV behavior of the partition function will be studied. The product that appears in the light cone partition function $z_{lc} (\tau^{\prime},\tau_0 )$  is a  massive generalization of the Theta functions. The modular properties of this kind of "generalized"  Theta functions were  studied in \cite{PandoZayas:2002hh,Greene:2002cd,Takayanagi:2002pi}. Here, the thermalization time $\tau_0$ plays a role in the modular transformations.\footnote{ For the time-independent case, in references\cite{Greene:2002cd,Takayanagi:2002pi} , the modular properties are studied keeping m independent of $\tau_2$, while  in \cite{PandoZayas:2002hh}, the dependence of m on $\tau_2$ is taken into account.} Consider the following generalized modular function
 \be
Z_{a,b}(\tau^\prime,\tau_0)=
\left |\prod_{n=-\infty}^{\infty}(1-e^{-2\pi\tau_2\sqrt{m^2(\tau_0)+(n+b)^2}
+2\pi i\tau_1(n+b)+2\pi ia}) \right |^8
\label{zetaab}
\ee
such that $z_{lc} (\tau^\prime,\tau_0 )$ can be written as
\be z_{lc}(\tau^\prime,t_0)= \frac{Z_{1/2,0}(\tau^\prime,\tau_0)}{Z_{0,0}(\tau^\prime,\tau_0)}
\ee  
Following  a similar strategy used in \cite{Greene:2002cd,Takayanagi:2002pi},which in fact consists of the same steps developed in (\ref{lnz}),(\ref{identi}) and (\ref{fstep}),  together with the Poisson resummation formula, it can be shown that
 
 \beq 
\frac{\ln z_{lc}(\tau^{\prime},\tau_0)}{8}&=&
\ln Z_{0,\frac{1}{2}}(\frac{\tau^\prime}{|\tau^{\prime}|^2},\tau_0-\ln |\tau^{\prime}|)
-\ln Z_{0,0}(\frac{\tau^\prime}{|\tau^\prime|^2},\tau_0-\ln |\tau^\prime|) \nonumber \\
&+&2\pi\frac{\tau_2}{|\tau^\prime|^2}\left[ 
\Delta_{\frac{1}{2}}(\frac{\tilde{\beta} fe^{-\tau_0}|\tau^\prime|}{2\pi\tau_2})
-\Delta_{0}(\frac{\tilde{\beta} fe^{-\tau_0}|\tau^\prime|}{2\pi\tau_2)}\right]
\label{pmodular}
\eea

where $\Delta_{1/2}(\frac{\beta fe^{-\tau_0}|\tau^\prime|}{2\pi\tau_2})$ and
$\Delta_{0}(\frac{\beta fe^{-\tau_0}|\tau|}{2\pi\tau_2})$  are defined by\footnote{Actually, $\Delta_b(t_0)$ corresponds to 
the zero-energy 
of a 2D massive complex scalar boson $\phi$ 
with twisted boundary condition 
$\phi(\tau,\sigma+\pi)=e^{2\pi ib}\phi(\tau,\sigma)$.}

\be
\Delta_b(m)=-\frac{1}{2\pi^2}\sum_{p=1}^{\infty}\cos(2\pi bp)
\int^{\infty}_{0} ds\ 
e^{-p^2s-\frac{\pi^2 m^2}{s}}=
-\frac{m}{\pi}\sum_{p=1}^{\infty}\frac{\cos(2\pi b p)}{p}
K_{1}\left(2 \pi mp\right)
\label{casimir}
\ee
and $K_{1}$ is a modified Bessel function of the second kind. 
Using (\ref{pmodular}) and setting $\tau_1=0$,  the leading behavior of $Z(\beta,\tau_0)$   as $\tau_2\rightarrow 0$ is
\be
exp \left\{-\frac{\tilde{\beta}^2}{2\pi\alpha'\tau_2}
+\frac{16\pi}{\tau_2}\left[ 
\Delta_{\frac{1}{2}}(\frac{\tilde{\beta} fe^{-\tau_0}|\tau|}{2\pi\tau_2})
-\Delta_{0}(\frac{\tilde{\beta} fe^{-\tau_0}|\tau|}{2\pi\tau_2})\right]\right\}
\ee
Thus, the partition function starts to diverge when the exponent above is zero. Hence,  the Hagedorn temperature satisfies the following equation
\be
\frac{\beta_H}{4\pi\alpha'}= 16\pi\left[ 
\Delta_{\frac{1}{2}}(\frac{\beta_H fe^{-\tau_0}}{2\pi\sqrt{2}})
-\Delta_{0}(\frac{\beta_H fe^{-\tau_0}}{2\pi\sqrt{2}})\right].
\label{hagedor}
\ee
 In the   asymptotically flat region: $fe^{-\tau_0}<<1$, on gets
\begin{equation}
    T_H=\frac{1}{2\pi\sqrt{2\alpha^\prime}}\left(1+2\sqrt{\alpha^\prime}fe^{-\tau_0}+...\right), \label{thflat}
\end{equation}
where dots mean higher order terms in the Ramond field. The first term is just the flat space result for the Hagedorn temperature. Note that as time goes from $\infty$ to $-\infty$, the Hagedorn temperature increases as the thermalization time goes towards the singularity. It can be shown that this happens at all instants of time and not necessarily in the approximation used in (\ref{thflat}).
  Let`s rewrite the difference that appears in (\ref{hagedor}) as
\be  
\Delta_{\frac{1}{2}}(\frac{\beta_H fe^{-\tau_0}}{2\pi\sqrt{2}})
-\Delta_{0}(\frac{\beta_H fe^{-\tau_0}}{2\pi\sqrt{2}}) = 
\frac{1}{2\pi^2}\sum_{p=1}^{\infty}\frac{[1-(-1)^p]}{p}\int^{\infty}_{0} ds\ 
e^{-p^2s-\frac{ \beta_H ^2f^2e^{-2\tau_0}}{8s}} ,
\ee
so, the derivative of $\beta_H$ with respect to $-\tau_0$ is  \be
- (2f\beta e^{-\tau_0})^2\left[\sum_{p=1}^{\infty}\frac{[1-(-1)^p]}{p}\int^{\infty}_{0} \frac{ ds}{s}\ 
e^{-p^2s-\frac{\beta_H ^2f^2e^{-2\tau_0}}{8s}}\right] < 0\ee 
and one concludes that the Hagedorn temperature $T_H=\frac{1}{\beta_H}$  increases as $\tau_0$ goes to $-\infty$, that is, as the string approaches the null singularity.

The Hagedorn behavior close to the null singularity can be  now cleared up just analyzing the asymptotic behavior  of $\Delta_b(x)$ defined in (\ref{casimir}).  This can be done using the method of Steepest Descents(see for example chapter  12 of \cite{Arfken}),or one just can use (\ref{besselint}) to represent $\Delta_{0}(\tau_0)$ and $\Delta_{1/2}(\tau_0)$ as modified Bessel functions and then use (\ref{besselassy}). As the higher values of p are exponentially suppressed,   the most  relevant term in the series (\ref{hagedor}) is given by taking $p=1$,

\beq
\lim_{\tau_0\rightarrow -\infty}\beta_H(\tau_0)&=&
\left[32\pi \alpha^\prime\sqrt{\frac{ f \sqrt{2}}{\pi}}\right]^{4/3}
\lim_{\tau_0\rightarrow -\infty}e^{-\frac{2\tau_0}{3}}\exp \left(-\frac{4\beta_H fe^{-\tau_0}}{3\sqrt{2}}\right)\nonumber \\
&=&0\,.
\label{aspphag}
\eea
So, as the singularity is approached, the Hagedorn temperature  is pushed toward infinity. It is tempting to conclude that there is no Hagedorn transition at any finite temperature close to singularity. 
This point will be brought again in the conclusions.

\section{Conclusions}

In the present work, the LvN formulation was used to define an invariant thermal density matrix for a superstring propagating in a  time dependent plane wave background with a constant self-dual Ramond-Ramond 5-form  and a linear dilaton in the light-like direction.  The metric has a cosmological singularity at $\tau \rightarrow - \infty$ and it  is asymptotically flat at $\tau \rightarrow\infty$.

In the formulation used here, it is assumed that the system enters thermodynamic equilibrium adiabatically at a certain time $\tau_0$. The adiabatic approximation is controlled by the Ramond field.  With this assumption, it was possible to use the density matrix to calculate the Hagedorn temperature as a function of $\tau_0$.  It has been shown that the Hagedorn temperature increases as string propagates from the asymptotically flat region  towards the singularity. In particular, the calculation shows that the Hagedorn temperature diverges at the singularity,which could indicate that, in this background, there is no hagedorn behavior near the singularity. However, we need to be careful in extrapolating the result found here to this region. This is because it is the region of strong coupling, owing to the dependence of the dilaton on time. It is important to keep in mind that the time that appears in the Hagedorn temperature is the time where thermalization occurs. So,in addition to the fact that a free string gas cannot be defined in this region, the very notion of thermodynamic equilibrium is not so simple to assume in the strong coupling limit. This is just because 
in non-interacting thermodynamics, one always starts with a weakly interacting gas and then the coupling is  adiabatically
turned down until to reach the free gas. If one doesn`t start with the interacting
gas the  equilibration processes cannot occur.  It is clear how to figure out  this process in the small dilaton region, but in the strong coupling limit it is not.  Note that the Ramond field also plays a role here, acting as the controller of adiabatic dynamics. 

On the other hand,  sometimes in string theory the perturbative string presents some window into the non-perturbative sector of the theory. This is the case for example of D branes seen as boundary states in the closed string channel. Perhaps, in view of \cite{Marchioro:2020qub}, \cite{Marchioro:2007ch} in this case the entanglement entropy may play this role. Let's clear up this point.  
It was shown in reference \cite{Marchioro:2020qub} that the left/right entanglement entropy is finite when evaluated at the singularity and thus it can be used to probe the null singularity. This entropy is actually the entropy related to the vacuum state as seen by asymptotic observers; this state is in effect a boundary state. Indeed, in reference \cite{Marchioro:2007ch}, for the time dependent pp wave studied in \cite{Papadopoulos:2002bg}, it was shown that the vacuum state, as seen by asymptotic observers, actually represents a D-brane described in the closed string channel. However, for that model, the dilaton remained small near the singularity. It would be interesting to verify this point for the model studied here. Also, it will be interesting to calculate the Hagedorn temperature as function of the dilaton. To this end, the finite temperature string field theory formulation developed in \cite{Abdalla:2005qs} will be extremely useful. This is a work in progress.

 Finally,the invariant density matrix also allowed to calculate the two-point thermal function in real time. It was shown that the real time thermal two point function can be written in terms of generalized Theta functions. The modular properties of these functions can be used to  study  non-equilibrium thermodynamic quantities, such as transport coefficients and the scrambling
time.  This will be left for a future work.


\bibliographystyle{JHEP}
\bibliography{biblio}
\end{document}